# Vibration control of a rotating cantilever beam using piezoelectric actuator and feedback linearization method

Roya Salehzadeh[1], Firooz Bakhtiari- Nejad[2], Mahnaz Shamshirsaz [3]

[1] Amirkabir University of Technology/Department of mechanical engineering; roya.salehzadeh@aut.ac.ir
[2] Amirkabir University of Technology / Department of mechanical engineering; baktiari@aut.ac.ir
[3] Amirkabir University of Technology / Department of mechanical engineering; shamshir@aut.ac.ir

**Abstract**

Vibration of various structures such as blades of turbines, helicopters and all kinds of rotating robot arms can damage the structures and disrupt their performance/balance. Thus, investigation of the reduction and control of vibration of these structures is significant. In this paper, the coupling and nonlinear flexural - torsional vibration of a rotating beam in the presence of piezoelectric layer is considered. The von Karman strain- displacements are used for deriving the potential equation. The nonlinear equations of motion are obtained based on Hamilton's principle and two ordinary differential equations are derived by the assumed mode method. Besides, a feedback linearization control strategy is proposed for suppressing the vibration of rotating beam.
According to the equation of motion, flexural and torsion vibration are coupled together. Therefore, control of flexural mode can result in control whole system.

**Introduction**

It is clear that rotary structures such as blades of turbine, helicopter and robot arms have various applications in many different fields of engineering. The simplest way to analyze vibration of the structures is considering it as a beam. Suppression of vibrations in rotating beams is noticed by researches, since vibration control in the structures have significant role of a kind that omitting noise, improving performance and decreasing of fatigue life. Therefore, modeling of dynamic, active vibration control and analysis of rotating beams stability are studied recently.
Many researchers have neglected the longitudinal vibration for vibration control of a rotating beam. Yang et al. [1] ignored the longitudinal vibrations of a rotating beam and controlled flexural vibrations by utilizing positive position feedback control and the momentum exchange feedback. Krenk et al. [2] studied the control of resonance vibrations of a rotating beam. Younesian and Ismail-Zadeh [3] presented a new method. They controlled flexural vibrations of rotating beam by using the increasing internal tensile force. Meng et al. [4] has been studied the vibration control of a rotating beam in Electro- rheological sandwich structures. They show that the vibration of the rheological rotating beam with different velocity and acceleration disappears quickly by applying electric field to the beam. Among the various methods of control, sliding mode method is more common to eliminate vibrations of a rotating beam. This method has been studied in [5], [6], and [7]. This method is a proper way to control systems with uncertainty [8]. Xue and Tang [5] studied the flexural vibration of a rotating beam and derived equations of motion without considering longitudinal vibration. They ignored from the longitudinal vibrations. As a result, non-linear coupling terms between longitudinal and flexural vibrations has been ignored. by considering the effects of nonlinear coupling between the hub circulating and flexural vibrations of beam, they designed a sliding mode controller to suppression the vibrations of the beam and demonstrated that the controller design is continuous and does not have chattering problems. Huang et al. [6] controlled the vibrations of a rotating flexible structure with considering all vibrational coupling terms, by using the finite element method and sliding mode control. Some actuators may be used for creation a control force to eliminate vibration. Since piezoelectric materials can be used as actuators and sensors, the piezoelectric transducer is used in control systems. For nonlinear vibration control of a rotating beam, piezoelectric transducers can be used in various forms [5], [6], [9]. These materials can be used as full layers or small pieces on the surface of structures. Lin [9] has used two layer of piezoelectric as actuator and sensor to control the vibration of a rotating beam with elastic foundation. Also, he used derivative- proportional controller to control the flexural vibrations of rotating beam. It should be noted that one of the problems of using a full layer of piezoelectric is that this kind of piezoelectric effect on the dynamic behavior of the system and increase the natural frequencies of the structures. Therefore, researchers study to obtain suitable dimension and position for piezoelectric pieces on the structure. For example, Huang et al. [6] used a piezoelectric piece and controlled vibration of rotating beam by using finite element method and the sliding mode controller. Also, Xue and Tang [5] by using a piece of piezoelectric and sliding mode technique controlled nonlinear flexural vibrations of a rotating beam. In this paper, the gyroscopic effect produced in the beam due to the base rotation is investigated. These gyroscopic terms in fact produce the coupling between flexural and torsional modes. If a cantilever beam undergoes flexural vibrations, and if the base of the beam undergoes rotational motion, torsional vibrations are induced in the beam due to Coriolis Effect. [10-12]. the magnitude of this coupling depends on the angular velocity of the base rotation.

In this paper, nonlinear coupling flexural- torsional vibrations of a rotating beam with piezoelectric layer is studied. Based on Hamilton's principle, equations of motion are obtained with considering displacement components in both vertical and torsional directions. Subsequently, these equations are discretized by using Assume mode method. Finally, a feedback linearization controller is designed for reducing vibration of the system. It is noticeable to mention that the voltage of the piezoelectric is considered as the control input.
The designed controller can control and decline flexural-torsional vibrations.

**Mathematical Modeling**

In this section, two nonlinear and coupled equations of motions of a rotating beam are developed. First, the assumptions and general modeling of the cantilever beam are presented. A uniform and initially straight cantilever beam with a layer of piezoelectric material on top of its surface as depicted in Figure. 1 is considered. Width of the piezoelectric layer is the same as the beam width. Moreover, the beam follows the Euler-Bernoulli beam theory; therefore shear deformation and rotary inertia are negligible.

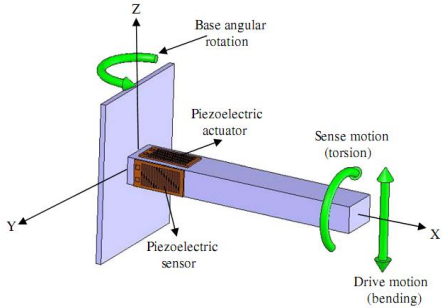

Figure 1: schematic of flexural- torsional vibrating beam [12]

First step is deriving the kinetic energy of system. Point $P$ on the neutral axis is moved to point $P^*$ [12].

$$r_{P^*} = r_P + u_{P^*} \, , \, r_P = s e_x \, , \, u_{P^*} = u e_x + v e_y + w e_z$$
$$\frac{d(r_{P^*})}{dt} = \frac{d(u_{P^*})}{dt} + \Omega \times (r_P + u_{P^*}) \quad (1)$$

In equation (1) $\Omega$ is the base rotation vector. In this paper is considered rotation only about z axis and the beam has no lateral and axial vibrations, by simplifying equation (1) the velocity of point $P^*$ can be reduced to:

$$\frac{d(r_{P^*})}{dt} = \dot{w} e_y - \Omega(s+u) e_z \quad (2)$$

For a rotating beam the translation kinetic energy can be expressed as following form:

$$T_1 = \frac{1}{2} \int_0^l \rho_b \left( \left( \frac{\partial^2 w}{\partial x^2} \right)^2 + s^2 \Omega^2 \right) dx \quad (3)$$

In general, each cross- section of the beam experiences an elastic displacement of its neutral axis and rotation. The rotation of the neutral axis, from the undeformed to the deformed position, is described using counterclockwise Euler- angle rotations with the angle of rotations denoted as shown in Figure 3.

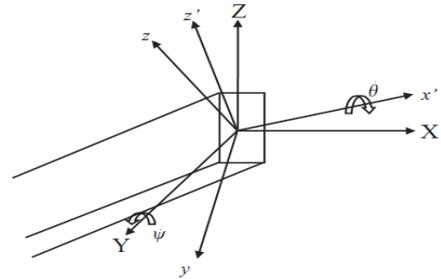

Figure 2: Euler angle rotations of the beam cross-section [12]

The transformation matrices can be written as:

$$\begin{Bmatrix} e_x \\ e_y \\ e_z \end{Bmatrix} = [R_\theta] \begin{Bmatrix} e_{x'} \\ e_{y'} \\ e_{z'} \end{Bmatrix} = [R_\theta][R_\psi] \begin{Bmatrix} e_\xi \\ e_\theta \\ e_\zeta \end{Bmatrix} = [R] \begin{Bmatrix} e_X \\ e_Y \\ e_Z \end{Bmatrix} \quad (4)$$

$$[R_\theta] = \begin{pmatrix} 1 & 0 & 0 \\ 0 & \cos\theta & \sin\theta \\ 0 & -\sin\theta & \cos\theta \end{pmatrix}, \, [R_\psi] = \begin{pmatrix} \cos\psi & 0 & -\sin\psi \\ 0 & 1 & 0 \\ \sin\psi & 0 & \cos\psi \end{pmatrix} \quad (5)$$

Consequently, undeformed and deformed coordinated are related together in the following manner [12]:

$$\begin{bmatrix} e_x \\ e_y \\ e_z \end{bmatrix} = \begin{pmatrix} \cos\psi & 0 & -\sin\psi \\ \sin\theta\sin\psi & \cos\theta & \sin\theta\cos\psi \\ \cos\theta\cos\psi & -\sin\psi & \cos\theta\cos\psi \end{pmatrix} \begin{bmatrix} e_X \\ e_Y \\ e_Z \end{bmatrix} \quad (6)$$

The angular velocity of system is obtained is given as:
$$\vec{\omega} = \Omega e_Z + \frac{\partial \psi}{\partial t} e_Y + \frac{\partial \theta}{\partial t} e_x \quad (7)$$

By substituting equation (6) into equation (7), the absolute angular velocity can be written as:
$$\vec{\omega} = \left( \frac{\partial \theta}{\partial t} - \Omega \sin\psi \right) e_x + \left( \frac{\partial \psi}{\partial t} \cos\theta + \Omega \sin\theta \cos\psi \right) e_y$$
$$+ \left( -\frac{\partial \psi}{\partial t} \sin\theta + \Omega \cos\theta \cos\psi \right) e_z \quad (8)$$

In this case is assumed that the angles of bending and torsion are small; therefore the expressions of the absolute angular velocity of the beam can be given as:

$$\vec{\omega}_x = (\frac{\partial \theta}{\partial t} - \Omega\psi)e_x, \vec{\omega}_y = (\frac{\partial \psi}{\partial t} + \Omega\theta)e_y, \vec{\omega}_z = \Omega s\, e_z \quad (9)$$

The rotational kinetic energy of system is given as follows:

$$T_2 = \frac{1}{2}\int_0^l \{I_{xb}\omega_x^2 + I_{yb}\omega_y^2 + I_{zb}\omega_z^2\}dx \quad (10)$$

In this paper, for the sake of simplicity, the effect of rotary inertia terms $(I_{yb}, I_{zb})$ is ignored. Therefore, the total kinetic energy is obtained as:

$$T = \frac{1}{2}\int_0^l \{\rho_b((\frac{\partial^2 \Omega}{\partial u^2})^2 + s^2\Omega^2) + I_{xb}\omega_x^2\}dx \quad (11)$$

The von Karman strain relation without considering axial vibration is given by:

$$\varepsilon_x = -z\frac{\partial^2 w}{\partial x^2} + \frac{1}{2}(\frac{\partial w}{\partial x})^2 \quad (12)$$

Equation (2) is valid if z is measured from the neutral surface of the beam. Since piezoelectric patch is not attached on the entire length of the beam, the neutral surface changes for each section. Wherever piezoelectric is not attached on the beam (x< l1 or x> l2), the neutral surface is the geometric center of the beam (z=0). For the portions where PZT is attached (l1<x<l2), the neutral surface, z, can be calculated as [13]:

$$\varepsilon_x = \begin{cases} -z\frac{\partial^2 w}{\partial x^2} + \frac{1}{2}(\frac{\partial w}{\partial x})^2 & \text{for } x\langle l_1 \text{ or } x\rangle l_2 \\ -(z-z_n)\frac{\partial^2 w}{\partial x^2} + \frac{1}{2}(\frac{\partial w}{\partial x})^2 & \text{for } l_1\langle x\langle l_2 \end{cases} \quad (13)$$

Therefore, the total potential energy of the system can also be expressed as:

$$U = \frac{1}{2}\int_0^L\int_A \left[ E\left(-z\frac{\partial^2 w}{\partial x^2} + \frac{1}{2}(\frac{\partial w}{\partial x})^2\right)^2 + G\left(\frac{\partial \theta}{\partial x}\right)^2 \right]dAdx \quad (14)$$

After simplification, we have:

$$U = \frac{1}{2}\int_0^L \left[ EI_y(x)\left(\frac{\partial^2 w}{\partial x^2}\right)^2 + \frac{1}{2}EA(x)\left(\frac{\partial w}{\partial x}\right)^4 + GJ(x)\left(\frac{\partial \theta}{\partial x}\right)^2 \right]dx \quad (15)$$

Where:
$\rho(x) = (\rho_b + S(x)\rho_p), \quad I_x(x) = (I_{xb} + S(x)I_{xp})$
$EI_y(x) = (EI_b + S(x)EI_p), \quad GJ(x) = (GJ_b + S(x)GJ_p)$
$EI_b = \frac{E_b w_b t_b^3}{12}, \quad EI_b = w_p E_p t_p^3\left[\frac{t_b^3}{4} + \frac{t_b t_p}{2} + \frac{t_p^2}{3} - \frac{z_n}{2}(t_b + t_p)\right]$
$EA(x) = (EA_b + S(x)EA_p), \quad S(x) = H(x-l_1) - H(x-l_2)$

(16)

And the virtual work expression of the system is in the following form:

$$\delta W_{nc} = \frac{1}{2}\int_0^L \frac{\partial^2 M_p}{\partial x^2}\delta w\, dx + C_B\int_0^L \frac{\partial w}{\partial t}\delta w\, dx + C_T\int_0^L \frac{\partial \theta}{\partial t}\delta\theta\, dx \quad (17)$$

Where:

$$M_p = -\frac{1}{2}bE_p d_{31}(t_b + t_p)V_p(t)S(x) = M_{p0}V_p(t)S(x) \quad (18)$$

And $C_B$, $C_T$ respectively are bending and torsion viscous damping coefficients.

**Governing equations of motion**
In this part, the Hamilton's principle will be used for deriving the equations of motions.

$$\int_{t_1}^{t_2}\{\delta T - \delta V + \delta W_{nc}\}dt = 0 \quad (19)$$

Substituting equations (11), (15) and (16) into equation (19), the equations of motion and corresponding boundary condition can be derived as [12, 14]:

$$\rho(x)\frac{\partial^2 w}{\partial t^2} + C_B\frac{\partial w}{\partial t} - I_x(x)\Omega\left(\frac{\partial^2 \theta}{\partial t\partial x} + \Omega\frac{\partial^2 w}{\partial x^2}\right) + \frac{\partial^2}{\partial x^2}\left(EI_y(x)\left(\frac{\partial^2 w}{\partial x^2}\right)\right)$$
$$+ \frac{3}{2}EA(x)\left(\frac{\partial w}{\partial x}\right)^2\left(\frac{\partial^2 w}{\partial x^2}\right) = \frac{\partial^2 M_p}{\partial x^2} \quad (20)$$

$$I_x(x)\left(\frac{\partial^2 \theta}{\partial t^2} + \Omega\frac{\partial^2 w}{\partial t\partial x}\right) + C_T\frac{\partial \theta}{\partial t} - \frac{\partial}{\partial x}\left(GJ\left(\frac{\partial \theta}{\partial x}\right)\right) = 0 \quad (21)$$

$w|_{x=0} = 0, \quad \frac{\partial w}{\partial x}|_{x=0} = 0, \quad \theta|_{x=0} = 0$

$I_x(x)\left(\frac{\partial^2 \theta}{\partial t^2} + \Omega\frac{\partial^2 w}{\partial t\partial x}\right) + C_T\frac{\partial \theta}{\partial t} - \frac{\partial}{\partial x}\left(GJ\left(\frac{\partial \theta}{\partial x}\right)\right) = 0$

$EI_y(x)\left(\frac{\partial^2 w}{\partial x^2}\right)|_{x=L} = 0,$

$GJ(x)\left(\frac{\partial \theta}{\partial x}\right)|_{x=L} = 0,$  (22)

It can be seen from equation (20) and (21) that the two partial differential equations of motion are coupled together through the base rotation with angular speed of $\Omega$.

**Numerical simulation**
The assumed mode model is used to convert the original partial differential governing equations of motion to ordinary differential equations. Using of this method, the lateral displacement $w$ and torsional displacement $\theta$ are assumed as linear functions of assumed modes and generalized coordinates as:

$$w(x,t) = \sum_{j=1}^n \varphi_j(x)p_j(t),$$

$$\theta(x,t) = \sum_{j=1}^n \psi_j(x)q_j(t) \quad (23)$$

By substituting equation (23) into equation (20) and (21), two nonlinear and coupled ODE equations are obtained:

$$M_{1ij}\sum_{j=1}^{\infty}\ddot{p}_j(t)+C_{Bij}\sum_{j=1}^{\infty}\dot{p}_j(t)+C_{1ij}\sum_{j=1}^{\infty}\dot{q}_j(t)\Omega+(K_{1ij}+D_{1ij}\Omega^2)\sum_{j=1}^{\infty}p_j(t) \quad (24)$$

$$+G_{1ijkl}\sum_{j=1}^{\infty}\sum_{k=1}^{\infty}\sum_{l=1}^{\infty}p_j(t)p_k(t)p_l(t)=F_{1i}$$

$$M_{2ij}\sum_{j=1}^{\infty}\ddot{q}_j(t)+C_{Tij}\sum_{j=1}^{\infty}\dot{q}_j(t)+C_{2ij}\sum_{j=1}^{\infty}\dot{p}_j(t)\Omega+K_{1ij}q_j(t)=0 \quad (25)$$

$$M_{1ij}=\int_0^L\rho(x)\varphi_i(x)\varphi_j(x)dx,\ C_{1ij}=\int_0^L I_x(x)\varphi_i(x)\psi'_j(x)dx$$

$$K_{1ij}=\int_0^L EI(x)\varphi''_i(x)\varphi''_j(x)dx,\ D_{1ij}=\int_0^L I_x(x)\varphi_i(x)\varphi''_j(x)dx$$

$$G_{1ijkl}=\frac{3}{2}\int_0^L EA(x)\varphi'_i(x)\varphi'_j(x)\varphi'_k(x)\varphi''_l(x)dx$$

$$F_{1ij}=M_{p0}V_p(t)\left[\varphi'_i(l_2)-\varphi'_i(l_1)\right]$$

$$M_{2ij}=\int_0^L I_x(x)\psi_i(x)\psi_j(x)dx,\ C_{2ij}=\int_0^L I_x(x)\psi_i(x)\varphi'_j(x)dx$$

$$K_{2ij}=\int_0^L GJ(x)\psi'_i(x)\psi'_j(x)dx,\ i,j=1,2,...,n$$

$$C_{Bij}=2\zeta_{1i}\omega_{1j}\ \text{for}\ i=j\ \text{and}\ C_{Bij}=0\ \text{for}\ i\neq j \quad (26)$$
$$C_{Tij}=2\zeta_{2i}\omega_{2j}\ \text{for}\ i=j\ \text{and}\ C_{Tij}=0\ \text{for}\ i\neq j$$

$$p=\{p_1,p_2,...,p_n\}^T,\ q=\{q_1,q_2,...,q_n\}^T$$

**Controller Design**

The control objective is to design the piezoelectric input voltage v (t) to suppression the vibration of beam. The truncated n-mode description for equation (24) and (25) using $r=\begin{bmatrix}p_1 & p_2 & \cdots & p_n & q_1 & q_2 & \cdots & q_n\end{bmatrix}^T$ can now be presented in the following matrix form:

$$M\ddot{r}+C\dot{r}+Kr=Fv_p(t) \quad (27)$$

The state space representation of the system dynamics, using $x=\begin{bmatrix}r^T & \dot{r}^T\end{bmatrix}^T$ is:

$$\dot{x}=A(x)x+B(x)v_p(t),$$

$$A=\begin{bmatrix}0 & I \\ -M^{-1}K & -M^{-1}C\end{bmatrix},\ B=\begin{bmatrix}0 \\ M^{-1}F\end{bmatrix} \quad (28)$$

Where M, K, C and F are the mass, stiffness, damping and force whose elements are shown in equation (26). In this work feedback linearization control is designed which the tip displacement of beam to approach zero Feedback linearization can be used as a nonlinear design methodology. The basic idea is to first transform a nonlinear system into a (fully or partia.ly) linear system, and then use the well-known and powerful linear design techniques to Complete the control design. The approach has been used to solve a number of practical nonlinear control problems. It applies to important classes of nonlinear systems (so-called input-state linearizable or minimum-phase systems), and typically requires full state measurement. However, it does not guarantee robustness in the face of parameter uncertainty or disturbances [14]. In feedback linearization method, the nonlinearities of system can be canceled by chosen $v_p$ as [15]:

$$v_p=B(x)^{-1}(x)(v-A(x)) \quad (29)$$

Thus, the control law:

$$v=-k_0x-k_1\frac{dx}{dt}-k_2\frac{d^2x}{dt^2}-...-k_{n-1}\frac{d^{n-1}x}{dt^{n-1}} \quad (30)$$

Where the $k_i$ chosen so that the polynomial $p^n+k_{n-1}p^{n-1}+...+k_0=0$ has all its roots strictly in the left-half complex plane, leads to the exponentially stable dynamics:

$$\frac{d^nx}{dt^n}+k_{n-1}\frac{d^{n-1}x}{dt^{n-1}}+...+k_1\frac{dx}{dt}+k_0=0 \quad (31)$$

**Results and Discussion**

In order to show the effectiveness of the proposed controller, the flexible beam depicted in Figure 1 is considered with the system parameters listed in Table 1.

Table 1: physical parameters of system

| Properties | symbol | Value | Unit |
|---|---|---|---|
| Beam length | $L$ | 0.15 | $m$ |
| Beam thickness | $t_b$ | 0.8e-3 | $m$ |
| Beam width | $b$ | 1.5e-2 | $m$ |
| Beam density | $\rho_b$ | 3960 | $kg/m^3$ |
| Beam elastic modulus | $E$ | 70 | $GPa$ |
| Beam shear modulus | $G$ | 30 | $GPa$ |
| First flexural damping ratio | $\zeta_{11}$ | 1 | % |
| Second flexural damping ratio | $\zeta_{12}$ | 0.16 | % |
| First torsional damping ratio | $\zeta_{21}$ | 1 | % |
| Second torsional damping ratio | $\zeta_{22}$ | 0.33 | % |
| Base rotation | $\Omega$ | 20 | $rad/s$ |

In this paper is assumed that the amplitude of the higher modes of the beam are very small compared with the first ones, the system can be truncated with j equal 2. Results have been obtained for the control schemes described for various condition of the system.

First, the controller efficiency is testing by applying the initial condition to the system. Figures (3) – (6) shows the first two modes of flexural and torsional vibration response of system for the case without controller and with controller which implemented voltage on the beam.

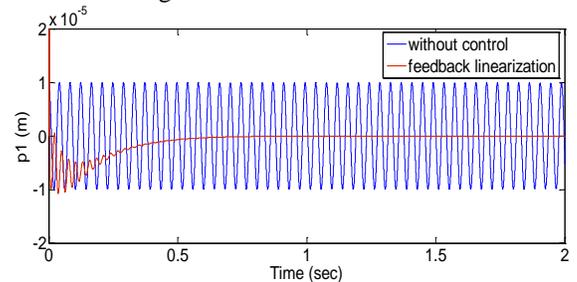

Figure 3: vibration response of the first mode of flexure

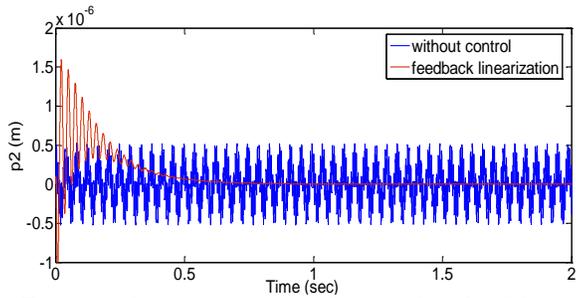
Figure 4: vibration response of the second mode of flexure

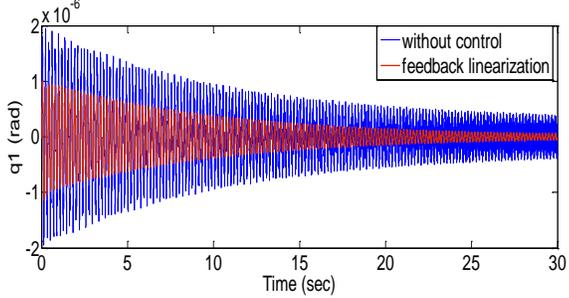
Figure 5: vibration response of the first mode of torsion

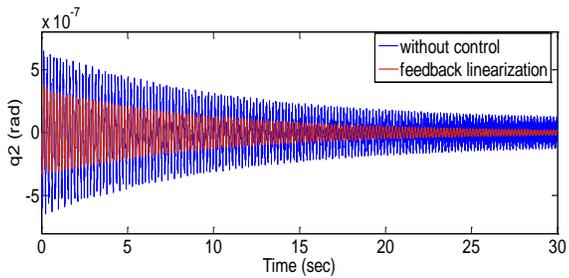
Figure 6: vibration response of the second mode of torsion

The tip displacements of the beam and control voltage are shown in Figures (7) and (8), respectively.

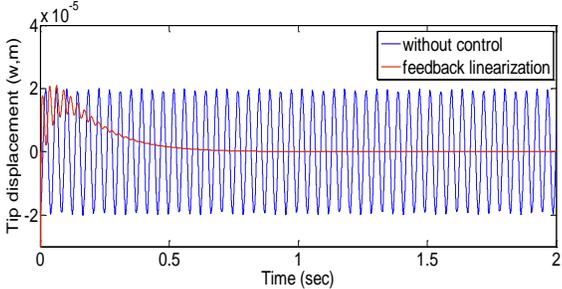
Figure 7: Tip deflection of the beam

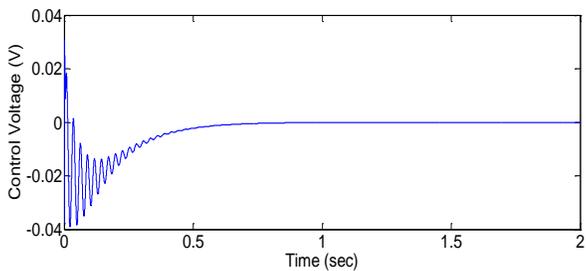
Figure 8: Control Voltage

It can be seen that the tip displacement and the vibrational response of the first two modes of flexure and torsion decay to zero.

In this part is considered harmonic disturbance for estimating performance of the controller. The amplitude and frequency are 0.001m and 24 Hz, respectively. That is applied to the first vibrational equation of flexural mode. The vibrational responses of the system for proposed controller with piezoelectric actuator are shown in Figures (9) - (12).

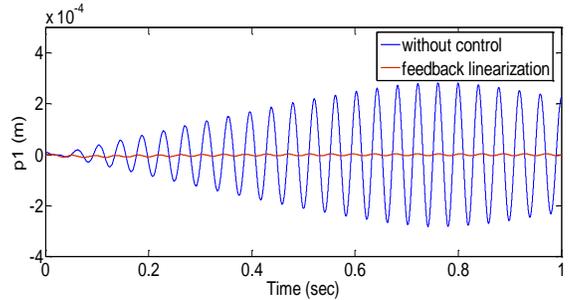
Figure 9: vibration response of the first mode of flexure

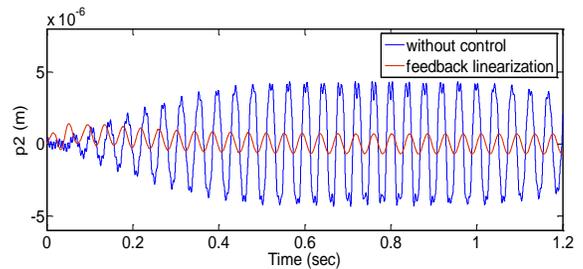
Figure 10: vibration response of the second mode of flexure

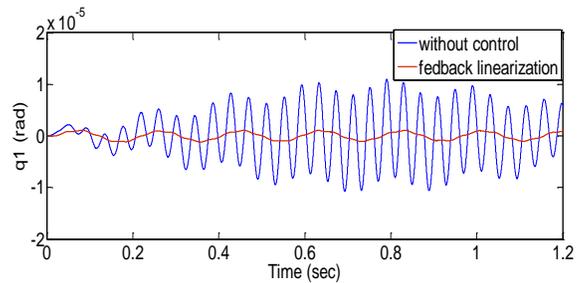
Figure 11: vibration response of the first mode of torsion

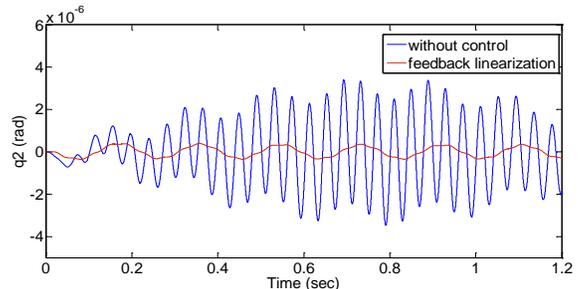
Figure 12: vibration response of the second mode of torsion

The results demonstrate that the beam vibration can be suppressed significantly using the piezoelectric actuator
  The tip displacement is shown in Figure (13).

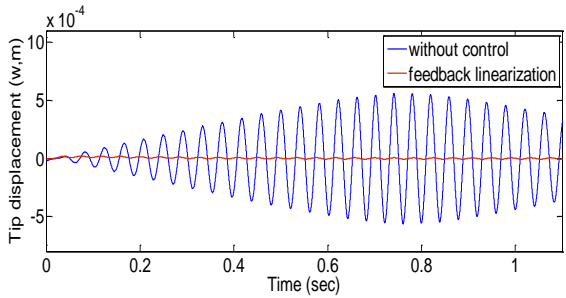

Figure 13: Tip deflection of the beam

Also the piezoelectric input voltage signal v (t) is depicted in Figure (14).

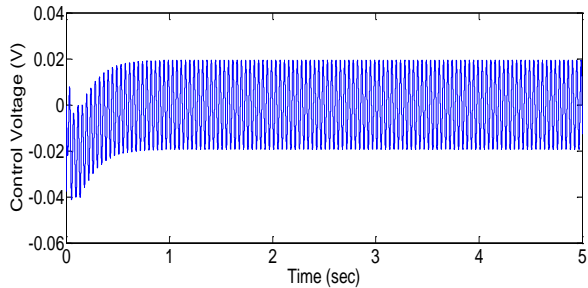

Figure 14: Control Voltage

**Conclusions**

In this paper, vibration of a beam with a piezoelectric actuator is analyzed and nonlinear coupled flexural and torsional equations are derived based on Hamilton's principle. An active feedback linearization controller is used for reducing the two first mode of flexural and torsional vibration of the rotating beam. The results reveal that modal function of the beam is well controlled under initial condition, harmonic disturbance. Therefore, the efficiency and accuracy of utilizing this controller to eliminate the vibration of the beam is proved.